# Crystal Growth and Basic Transport and Magnetic Properties of $MnBi_2Te_4$

Poonam Rani[1], Ankush Saxena[1,2], Rabia Sultana[1,2], Vipin Nagpal[3], S.S. Islam[4], S. Patnaik[3], and V.P.S. Awana[1,2,*]

[1] National Physical Laboratory (CSIR), Dr. K. S. Krishnan Road, New Delhi-110012, India

[2] Academy of Scientific and Innovative Research (AcSIR), Ghaziabad-201002, India

[3] School of Physical Sciences, Jawaharlal Nehru University, New Delhi-110067, India

[4] Centre for Nanoscience and Nanotechnology, Jamia Millia Islamia, New Delhi-110025, New Delhi, India

Abstract

We report successful growth of magnetic topological insulator (MTI) $MnBi_2Te_4$. The heating schedule basically deals with growth of the crystal from melt at $900^0$C and very slow cooling ($1^0$C/hr) to around $600^0$C with 24 hours hold time, followed by cooling to room temperature. Our detailed, PXRD Reitveld analysis showed that the resultant crystal is dominated mainly by $MnBi_2Te_4$ and minor phases of $Bi_2Te_3$ and MnTe. The transport measurements showed a step like behavior at around 150K followed by cusp like structure in resistivity at around 25K ($T_P$) due reported anti-ferromagnetic ordering of Mn. Both the resistivity transitions are seen clearly in dρ/dT measurements at 150K and 20K respectively. The 25K transition of the compound is also seen in magnetic susceptibility. Low temperature (5K) magneto-resistance (MR) in applied field of up to 6 Tesla exhibited –ve MR below 3 Tesla and +ve for higher fields. Also, seen are steps in MR below one Tesla. The studied $MnBi_2Te_4$ MTI crystal could be a possible candidate for Quantum Anomalous Hall (QAH) effect.

**[*]Corresponding Author**

Dr. V. P. S. Awana: E-mail: awana@nplindia.org
Ph. +91-11-45609357, Fax-+91-11-45609310
Homepage: awanavps.webs.com

Topological Insulators (TIs) are dominating the quantum condensed matter for over a decade by now [1-3]. In this direction, one of the most happening new advancement had been the magnetic topological insulators (MTIs). In case of MTIs, a magnetic layer or element is inserted among the running TI unit cells of bulk 3D topological insulators such as 3d metal doped $Bi_2Se_3$, $Bi_2Te_3$ and $Sb_2Te_3$ [4-6]. The insertion of magnetic layer along running 3D bulk topological insulators shifts the Dirac position and thus alters the quantum transport properties of the parent system [7-10]. One of the most fascinating properties of the MTIs is the appearance of Quantum Anomalous Hall (QAH) effect [7-9].

QAH happens due to the finite Hall Voltage created due to magnetic polarization and spin-orbit coupling, while the external magnetic field is absent. QAH is found to be in integer multiple of $e^2/h$ which is called Landau Level [9]. In principle, the inherent magnetism and topological electronic states are the necessary conditions for QAH. Though, ultra thin films of TIs with enhanced surface area may exhibit QAH due to strong spin orbit coupling (SOC), but is a rare possibility and if at all the same takes place at ultra low temperatures [10]. Other possibility to realize QAH is to dope the 3D bulk topological material with transition metal elements viz. Co, Cr, Eu etc. [11, 12]. In this case as well because of randomly distributed 3d metal impurity to the realization of QAH is often difficult and if at all is at mK temperatures [10-13]. The best solution till date to observe QAH at higher temperatures is via the MTIs. A magnetic topological insulator having strong SOC exhibits quantized resistance and non-dissipative current at room temperature [14]  In case of MTIs, a 3d metal based magnetically ordered layer is inserted between the running 3D bulk topological insulator. An ideal example, being put forward by theoreticians [15-17] and very recently realized by experimentalists is $MnBi_2Te_4$ [18-20]. $MnBi_2Te_4$ can be formulated as $Bi_2Te_3$ and MnTe, here the former is the 3D bulk topological insulator and later the magnetically ordered layer being inserted in the periodic structure at van der Waals gaps [21]. $MnBi_2Te_4$ is the first 3D antiferromagnetic topological insulator [18-20, 22]. Interestingly, there are only scant reports on $MnBi_2Te_4$, in particular the single crystals [22].

Keeping in view, the importance of QAH, the MTIs and the first 3D antiferromagnetic topological insulator, i.e. $MnBi_2Te_4$, we report here the crystal growth and characterization of the same. Detailed, PXRD Reitveld analysis showed that the resultant crystal is dominated by $MnBi_4Te_7$, $MnBi_6Te_{10}$ phases along with some MnTe and $Bi_2Te_3$ content. Resistivity measurements exhibited two transitions at 150K and 25K. The studied $MnBi_2Te_4$ MTI crystal could be a possible good candidate for Quantum Anomalous Hall Effect (QAHE).

The constituent elements Mn, Bi and Te of better than 3N purity are weighed (1gram) in stoichiometric ratio and grind in a glove box filled with Argon. Proper care is taken to avoid the exposure of the material from air. After grinding the powder is converted into pallet, which is further sealed into a vacuum tube under the pressure of $10^{-5}$mb. Now the sealed tube is held in the electronically controlled furnace at high temperature. The temperature goes up to 900˚C with heating rate of $120^0$C/hour and the temperature of furnace remains hold at 900˚C for 12hour. Now the temperature starts to decrease at the rate of 1˚C/hour and goes down to 600˚C. This temperature remains hold for 12hour followed by normal cooling to room temperature. The schematic of details of the heat treatment is given in inset (a) of Fig. 1. The powder X-ray diffraction pattern (PXRD) of the gently crushed part of crystal is taken on Rigaku X-ray diffractometer in the range 10˚ to 80˚ of 2θ˚ at scan rate of 2˚/min. Scanning Electron Microscope (SEM) make Bruker is used to visualize the morphology of the studied crystal. Raman spectrum of the crystal piece is taken on LabRam HR800-JY equipped with a laser source of 514nm. The spectrum is taken in wave number range of $50cm^{-1}$ to $400cm^{-1}$. Resistivity versus temperature measurements are done on a Physical Property Measurement System (PPMS) in temperature range of 300K down to 10K on a close cycle refrigerator.

Fig. 1 shows the powder XRD (X-ray diffraction) pattern of the part of crushed $MnBi_2Te_4$ crystal. The full proof Reitveld analysis is carried out on the observed pattern. The PXRD pattern is similar to that as observed recently for $MnBi_2Te_4$ self flux grown crystals [23]. The sample is found to be crystallized within space group R-3m and lattice parameters a = 4.3825(4)Å and c = c = 42.6849(1)Å. There are some un-reacted lines not fitting with the main pattern and these belong to $Bi_2Te_3$ shown in inset (a) of Fig. 1. The Wyckoff and coordinate positions obtained from Reitveld analysis are given in table 1. The situation is very similar to that as reported in ref. 23. The goodness of fitting parameters is $\chi 2$ = 3.05, which is reasonably good. The VISTA based software drawn unit cells for major $MnBi_2Te_4$ is given in inset (b) of Fig.1. The c-parameter for is around 40Å. The general formula for these homologous series of compounds can be simplified as follows MnTe + $nBi_2Te_3$, with n =1, 2, 3, as $MnBi_2Te_4$, $MnBi_4Te_7$ and $MnBi_6Te_{10}$. So, principally it depends upon the fact that after how many unit cells of $Bi_2Te_3$ an MnTe magnetic layer is inserted [23]. In present case, the PXRD of our crushed crystal showed mainly the $MnBi_2Te_4$ phase along with minor content of $Bi_2Te_3$ and MnTe. To elaborate further the expanded part of the fitted and observed PXRD is shown in inset (a) of Fig. 1. In earlier attempt, we tried to fit the PXRD data in MnTe + $nBi_2Te_3$, with n =1, 2, 3, as $MnBi_2Te_4$, $MnBi_4Te_7$ and $MnBi_6Te_{10}$ phases as well and the main phase was found to be distorted $MnBi_2Te_4$ i.e., $MnBi_6Te_{10}$ (56%) with space group R-3m, followed by, $MnBi_2Te_4$ along with $MnBi_4Te_7$ [24].

Fig.2 depicts the scanning electron microscope (SEM) picture of the surface of the as grown crystal. The slab like layered growth can be visualized and the uni-directional layered structure is clearly visible. We also did the EDAX to know the composition of the crystal at various points on the micrograph. The resultant composition was although near stoichiometric for Bi and Te but less in Mn content. The Raman spectrum of the as grown $MnBi_2Te_4$ taken at Laser wavelength of 514nm is given in Fig. 3. A broad peak with three to four main shoulders is seen in the range of 50 to $200cm^{-1}$. We de-convoluted the main peak being at $120cm^{-1}$ consisting of $102cm^{-1}$ and $140cm^{-1}$shoulders. The main peak at 120nm is having maximum intensity and two others at lower and higher wave numbers show much lesser intensity. The general overall look of the Raman spectrum shown in Fig. 3 is similar to that as reported in ref. 23 for $MnBi_2Te_4$.

The resistivity (ρ) versus temperature (T) plot for studied $MnBi_2Te_4$ crystal is given in Fig.4 in temperature range of 300K down to 10K, which is metallic in nature with a step like change in slope at around 150K ($T_P$), followed by a small kink (inset Fig. 4) at around 25K ($T_N$). The 25K transition is very feeble and hence for more clarity is shown as the derivative of resistivity (dρ/dT) plot against temperature (T) in Fig. 5. The inset of Fig. 5, clearly marks the 25K ($T_N$). Both $T_P$ and $T_N$ are clearly seen in the dρ/dT versus T plot and are marked in the plot. The 25K transition in ρ(T) of $MnBi_2Te_4$ is reported earlier [15-18, 22] and is due to the Anti ferromagnetic ordering of Mn spins [18, 22] in MnTe layers of running $Bi_2Te_3$ 3D topological insulator. To elaborate more on the transport behavior, Here not only the 150K transition of metallic resistivity is seen, but the Mn magnetic ordering related 25K transition is also seen clearly. Though, magnetic transition in the range of 20 to 30K is known to be due to Mn ordering [15-18, 22-27], the origin of 150K transition is not clear as of now.

The magnetic moment (M) versus temperature plot in temperature range of 50 to 10K in both field cooled (FC) and Zero Field Cooled (ZFC) situations at 0.1 Tesla for studied $MnBi_2Te_4$ is depicted in Fig. 6. An anomaly is seen in M(T) plot at below 30K, and the same seen visibly in 1/M versus T plot in inset of Fig. 6. The anomaly in magnetic susceptibility occurs nearly at the same temperature as being seen in transport measurements in Fig.4 and Fig. 5.

Fig. 7 shows the magneto-resistance measurements done to 5K in applied magnetic field of up to 6Tesla. The magneto-resistance decreases with increase in field of up to say one Tesla ($H_1$) and for higher fields starts increasing and becomes positive above 3 Tesla ($H_2$). Clearly, the magnetic ordering of Mn in MnTe layer of $MnBi_2Te_4$ has an effect on the magneto transport of the same. The expanded region (below $H_1$) of resistivity against magnetic field for the studied crystal is shown in inset of Fig.7. It is clear that initially the resistivity versus field plot is linear but starts showing step above say 0.5 Tesla, which are reminiscent of quantum transport. Detailed experiments for observation of Quantum Anomalous Hall (QAH) effect are underway in the magnetic field regime of much less than $H_1$ in different protocols.

In summary, the short communications reports the growth and basic physical properties of the magnetic topological insulator $MnBi_2Te_4$. Interestingly, the growth of topological insulators and so much so the newest one i.e. magnetic topological insulator viz. the presently studied $MnBi_2Te_4$ is not straight forward. Long heat treatments with slow cooling rates of up to $1^0$C/hour are required. Most of the literature on $MnBi_2Te_4$ single crystal growth and its characterization is too recent and is mostly on cond-mat arXiv [15, 19, 24, 25-32]. Keeping in view that $MnBi_2Te_4$ is an important new bulk 3D magnetic topological insulator hence its quality crystals are warranted, the present communication adds detailed growth parameters and basic electrical and magnetic characterization of the same to the scientific literature. We also synthesized the $MnBi_2Te_4$ crystal from higher temperature melt ($1000^0$C) as well and found that though the crystallized phase is same as in present case, the magneto-transport is slightly different albeit with similar features [33]. We conclude that present growth parameters are the optimum ones for obtaining the $MnBi_2Te_4$ magnetic topological insulator.

**Table 1:** Crystal structure data and details of data refinement for the investigated phases of as grown $MnBi_2Te_4$ crystal

| | Wyckoff | x | y | z |
|---|---|---|---|---|
| $MnBi_2Te_4$<br>Space group R-3m<br>a = 4.3825(4)<br>c = 42.6849(1) | | | | |
| Bi1 | 6c | 0 | 0 | 0.4304(1) |
| Te1 | 6c | 0 | 0 | 0.1429(5) |
| Te2 | 6c | 0 | 0 | 0.2873(6) |
| Mn1 | 3a | 0 | 0 | 0 |

**Figure Captions**

**Figure 1** Rietveld fitted XRD pattern of powder form of $MnBi_2Te_4$ crystal, inset (a) is the extended part of XRD and (b) the unit cell for the $MnBi_2Te_4$ crystals.

**Figure 2** SEM image of $MnBi_2Te_4$, depicting clearly the layered growth.

**Figure 3** Raman of $MnBi_2Te_4$ single crystals.

**Figure 4** Resistivity versus temperature plot of $MnBi_2Te_4$ Crystal, inset shows the zoom part of same, marking the possible 25K magnetic ordering of Mn.

**Figure 5** Derivative of resistivity versus temperature plot of $MnBi_2Te_4$ crystal, marking the possible 25K magnetic ordering of Mn, the inset shows the zoom part of the same.

**Figure 6** Magnetic moment (M) vs temperature (T) plot for $MnBi_2Te_4$ crystal in both Zero-Field-Cooled (ZFC) Field Cooled (FC) situations at 0.1Tesla, possible AFM ordering temperature ($T_N$) for Mn at below 30K is marked as a dip in M(T) plot, inset shows the 1/M vs T plot.

**Figure 7** Magneto resistances versus field plot for $MnBi_2Te_4$ crystal in applied magnetic field of up to 6 Tesla, inset shows the resistivity versus field plot for the same in lower applied fields of up to 1Tesla, marking the possible quantum steps.

Figure 1

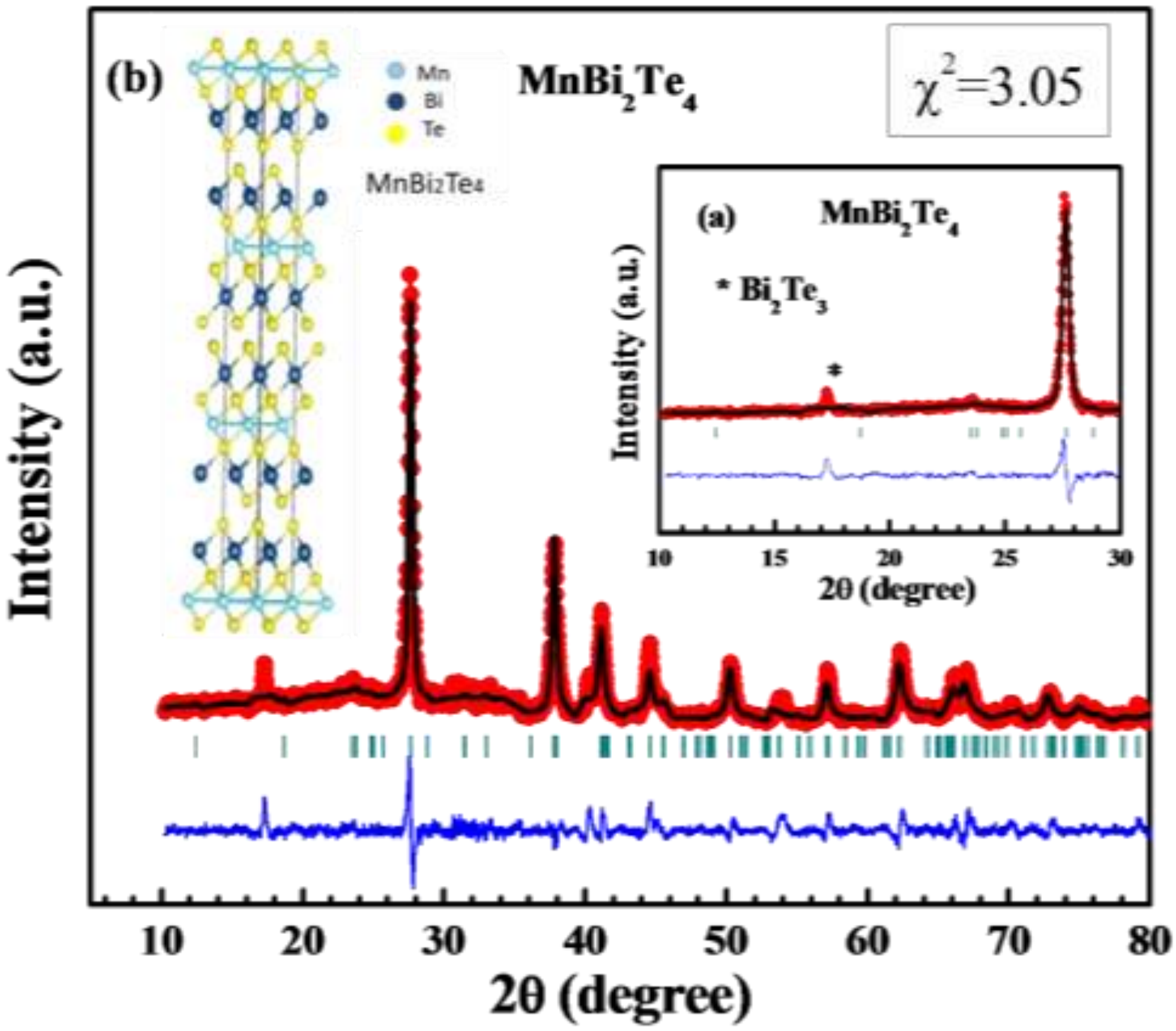

Figure 2

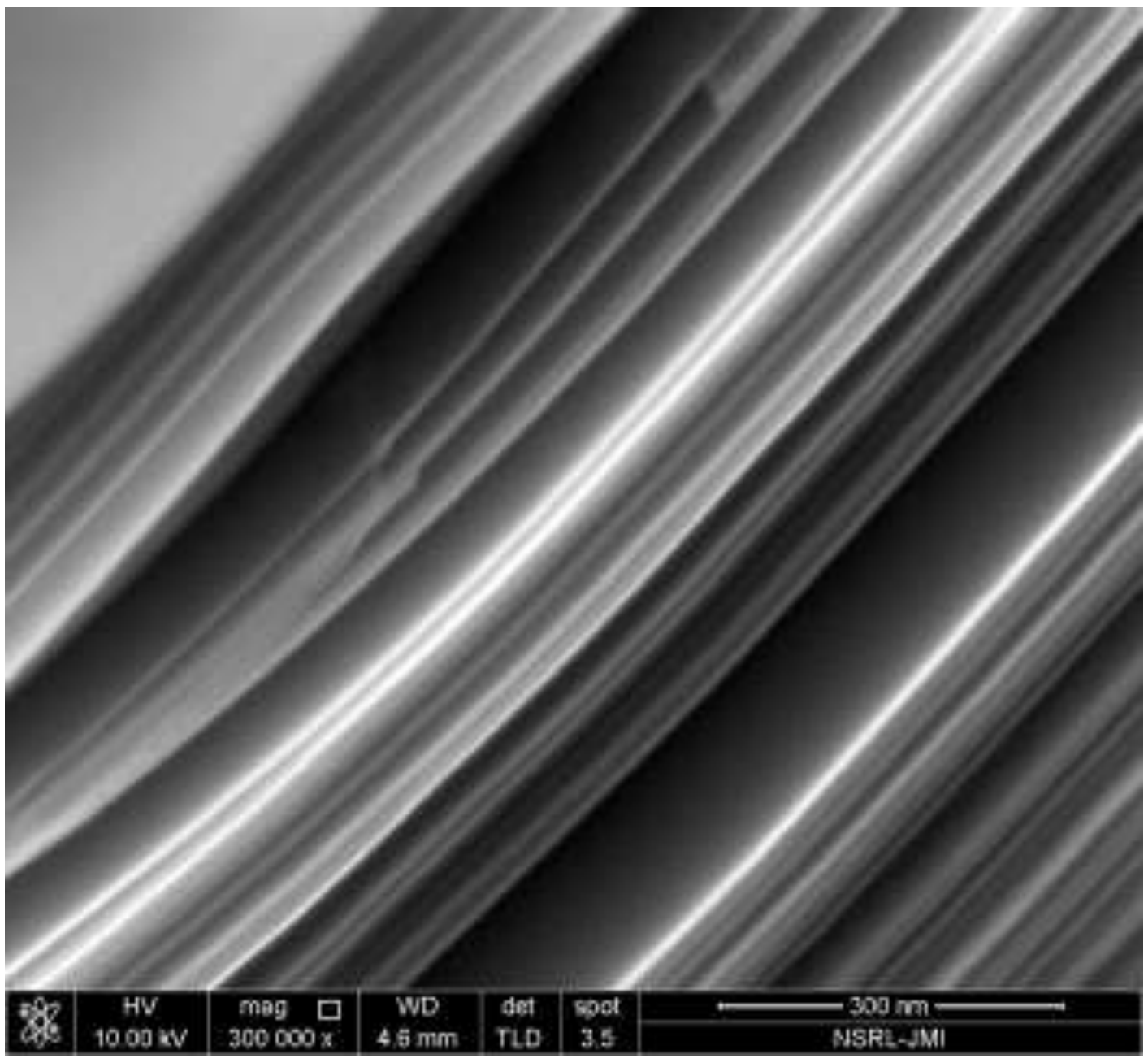

Figure 3

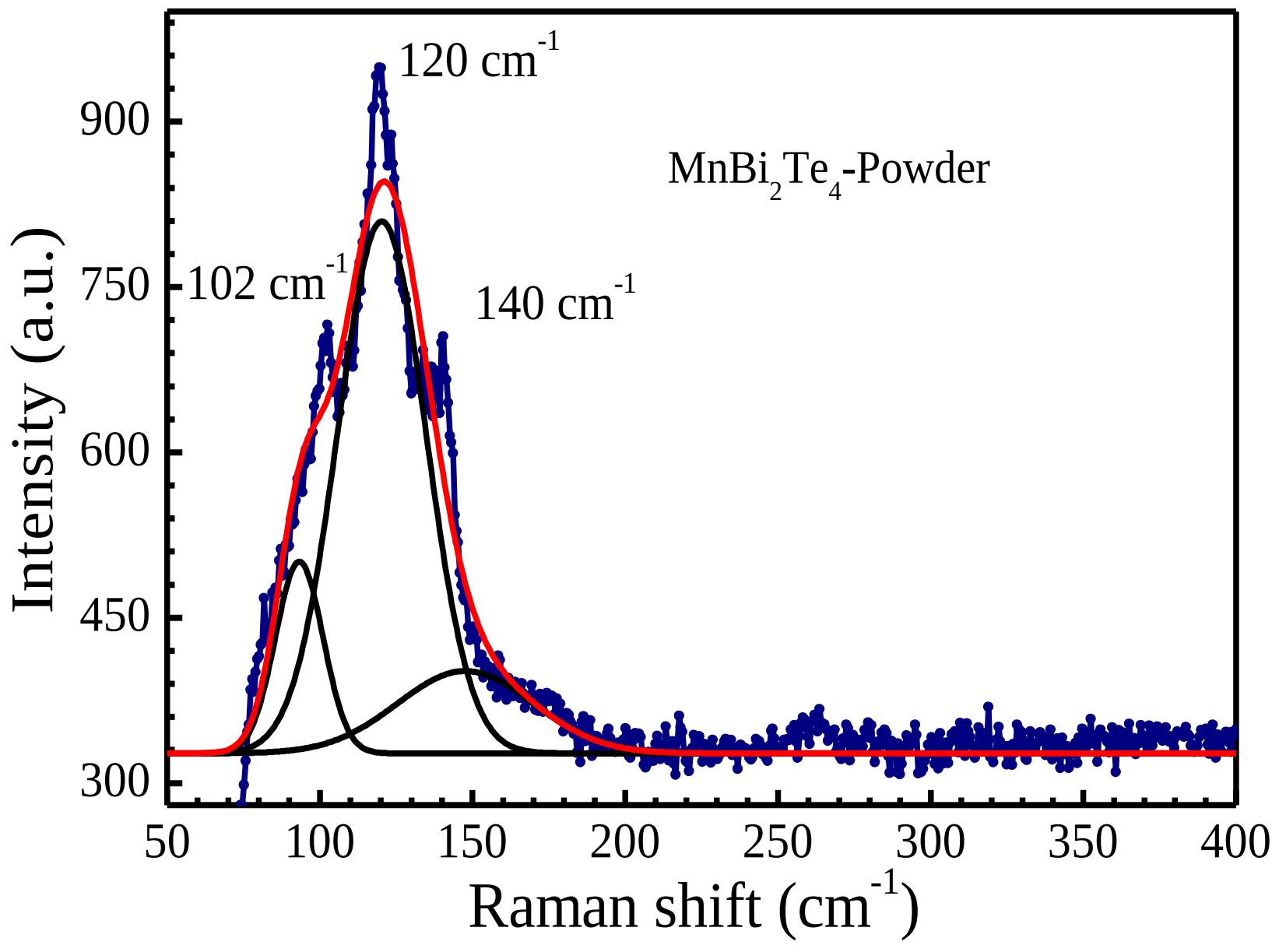

Figure 4

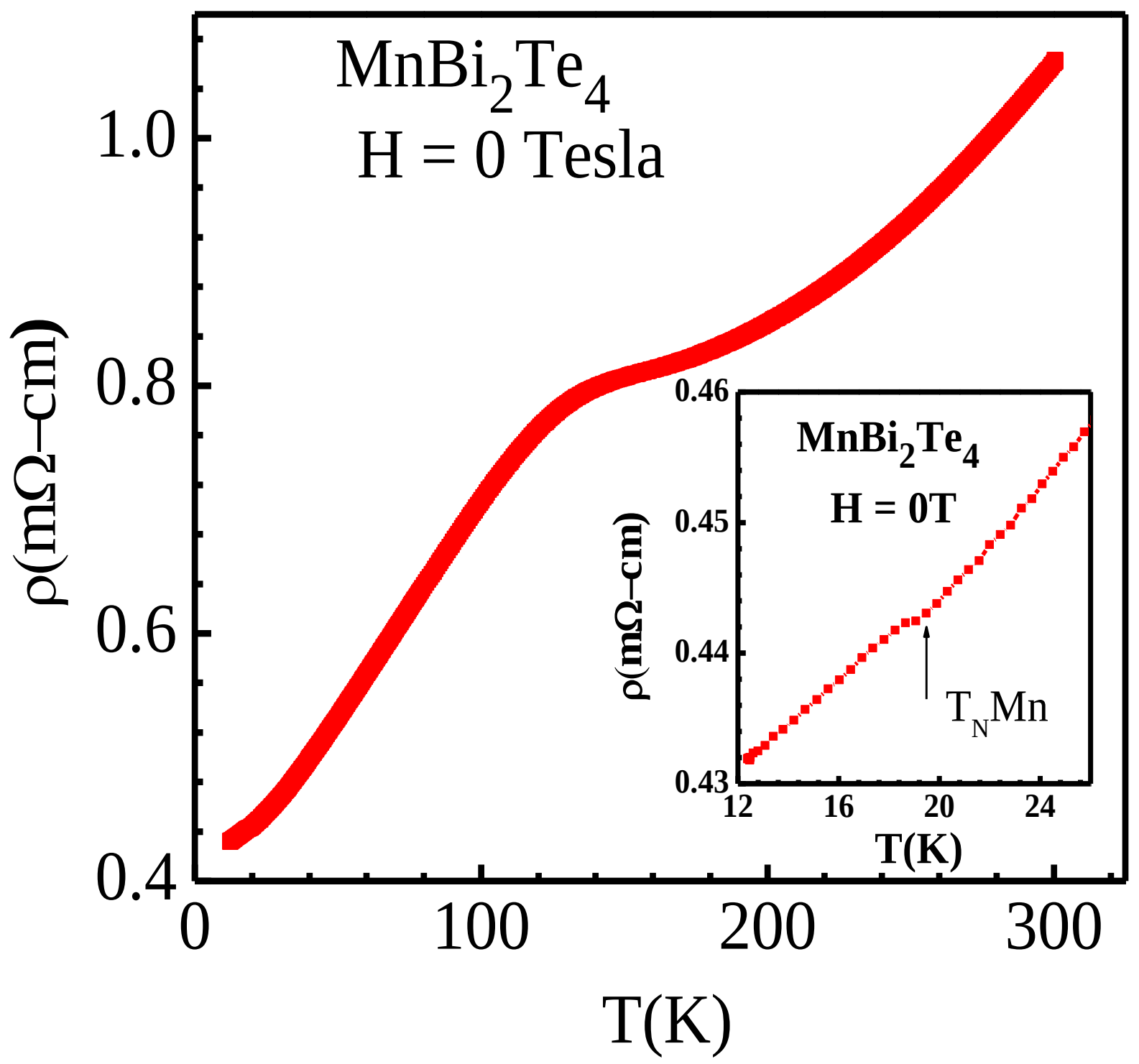

Figure 5

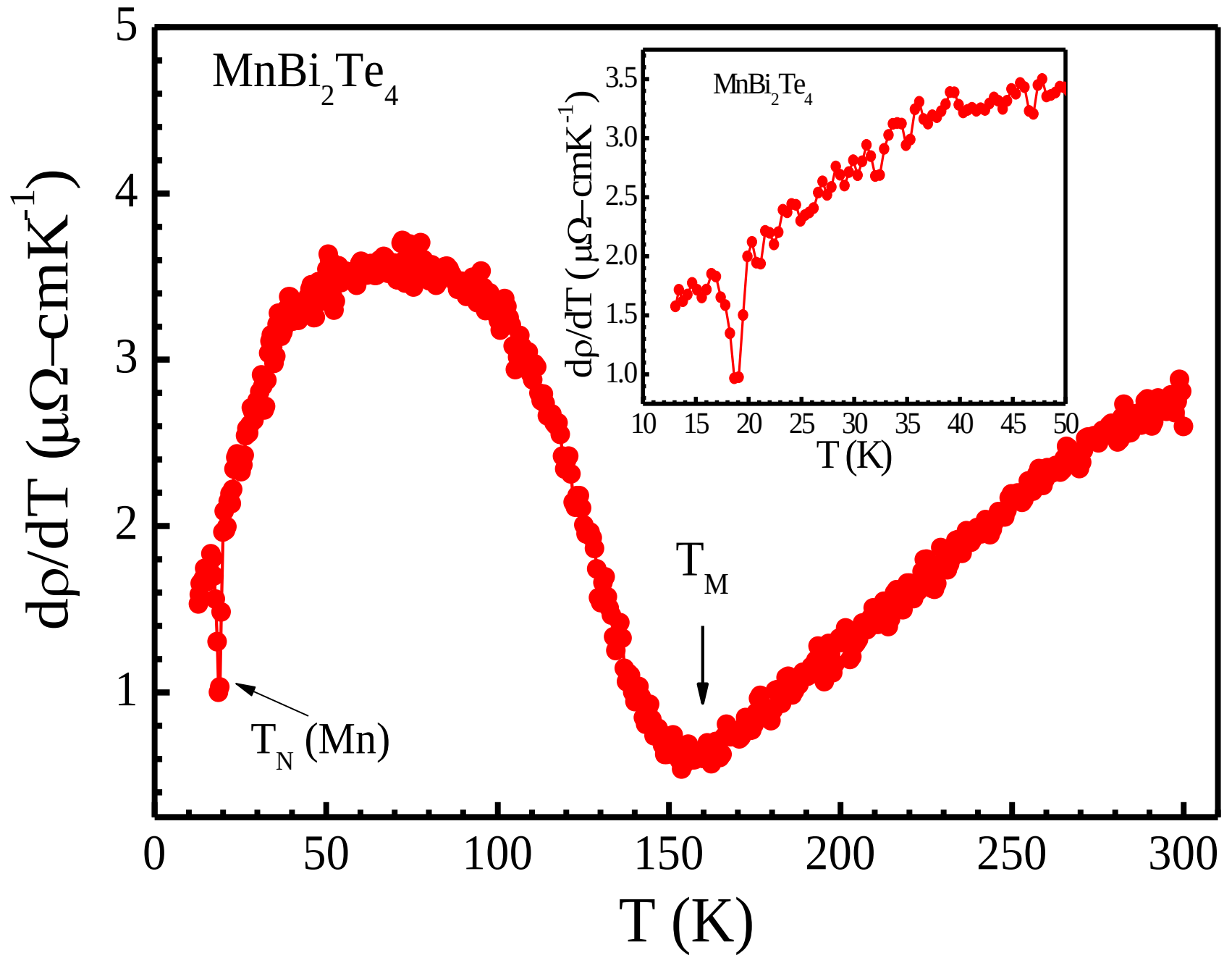

Figure 6

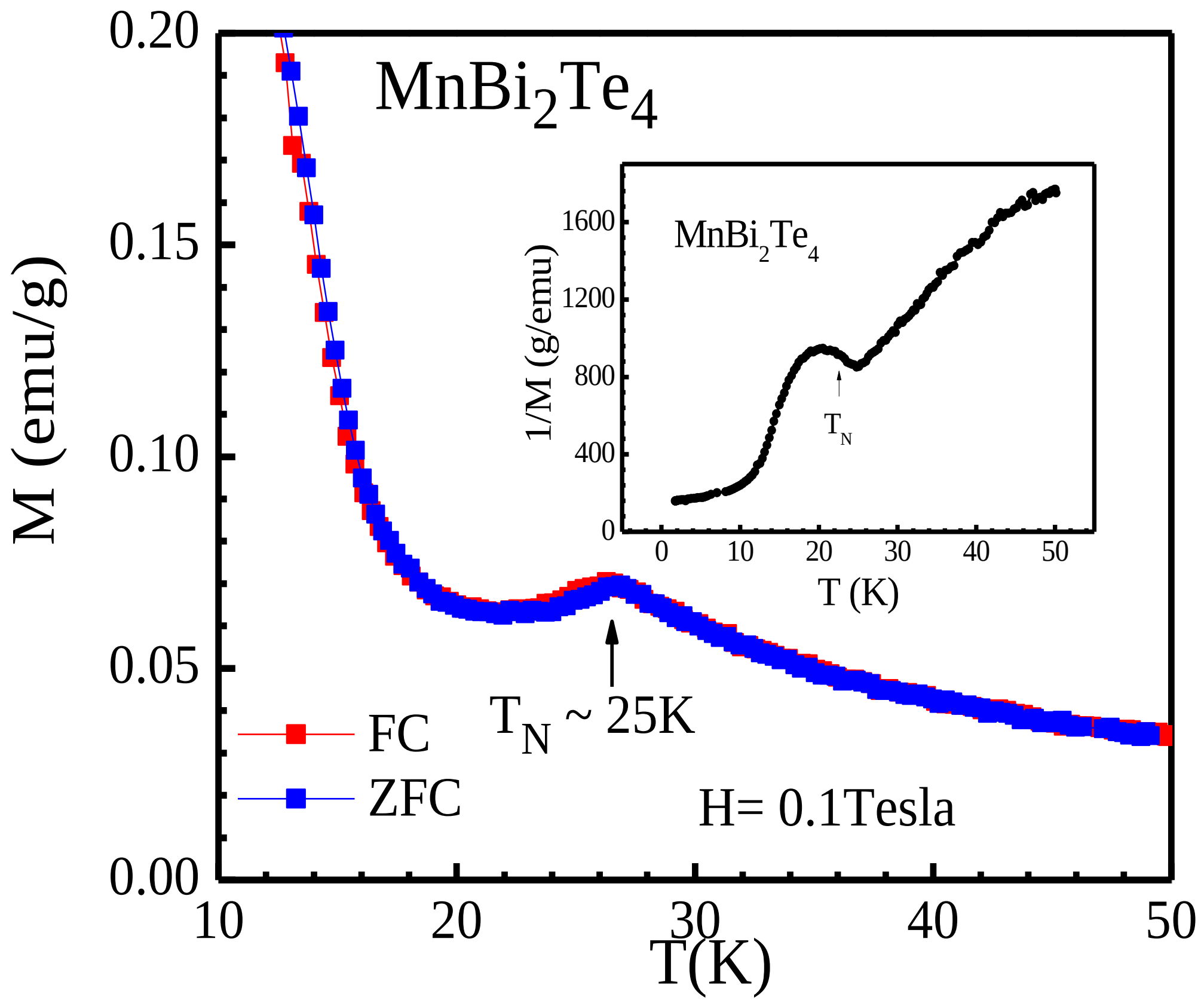

Figure 7

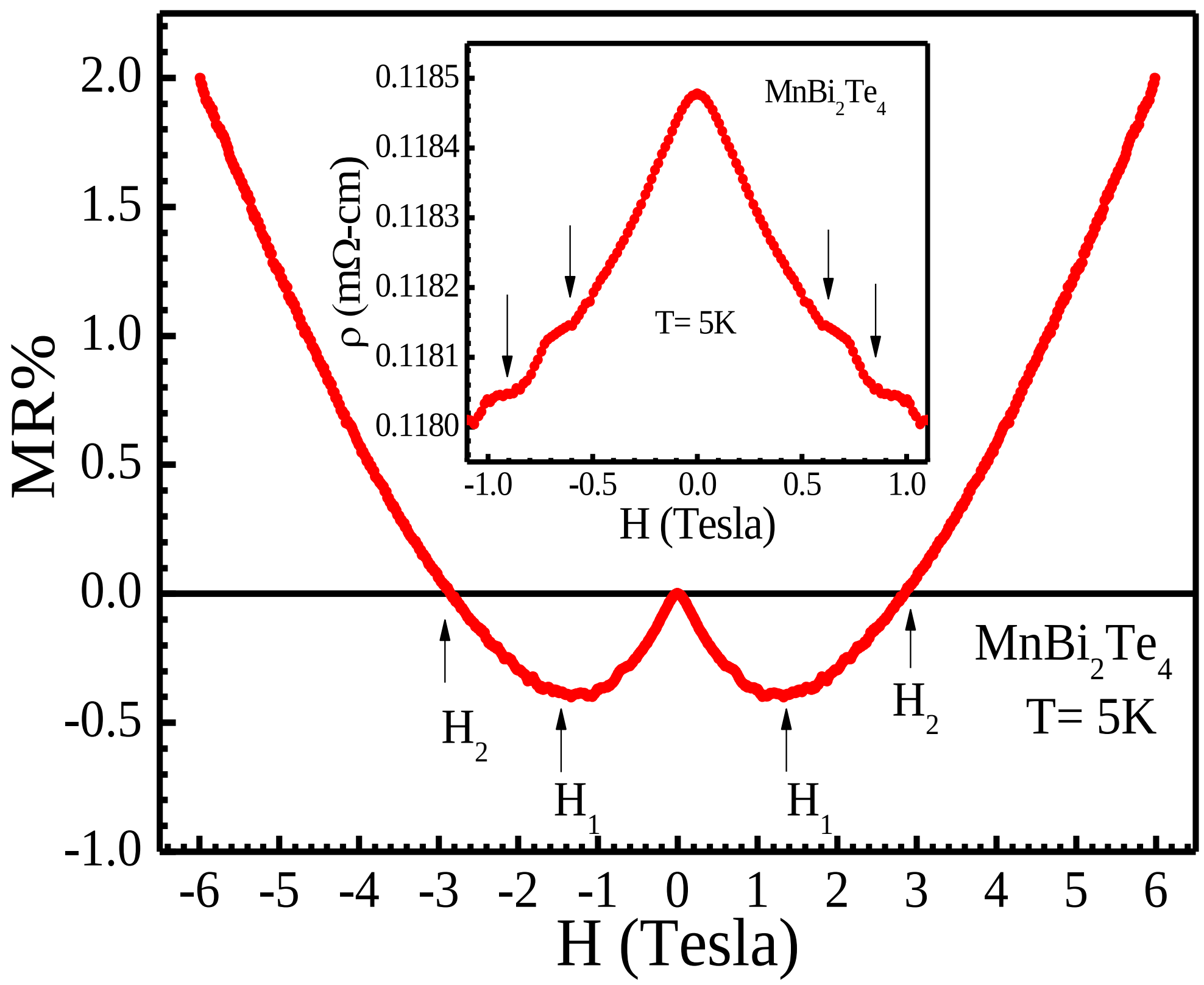

MR%
2.0
1.5
1.0
0.5
0.0
-0.5
-1.0
-6 -5 -4 -3 -2 -1 0 1 2 3 4 5 6
H (Tesla)
H2
H1
H1
H2
MnBi2Te4
T= 5K
ρ (mΩ-cm)
0.1185
0.1184
0.1183
0.1182
0.1181
0.1180
-1.0 -0.5 0.0 0.5 1.0
H (Tesla)
MnBi2Te4
T= 5K